\documentclass[twocolumn]{aastex631}
\usepackage{CJK}

\definecolor{mygray}{gray}{0.6}
\definecolor{magenta}{rgb}{0.858, 0.188, 0.478}
\definecolor{thp}{rgb}{0.455, 0.204, 0.506}
\usepackage{xspace}

\newcommand{\hjadd}[1]{{{#1}}}

\newcommand{\coadd}[1]{{{#1}}}
\newcommand{\coch}[2]{{{#2}}}
\newcommand{\hjch}[2]{{{#2}}}

\newcommand{\xxx}[1]{\textcolor{blue}{\textbf{xxx}\xspace}}

\newcommand{\fg}[1]{Fig.~\ref{fig:#1}}

\newcommand{\tb}[1]{Table~\ref{tab:#1}\xspace}
\newcommand{\se}[1]{Sect.~\ref{sec:#1}\xspace}
\newcommand{\Se}[1]{Section~\ref{sec:#1}\xspace}

\shorttitle{Spatial Correlation between Substructures}
\shortauthors{Jiang, Zhu, Ormel}
\graphicspath{{./}{figures/}}

\begin{document}
\begin{CJK*}{UTF8}{gbsn}

\title{No Significant Correlation between Line-emission and Continuum Substructures in the Molecules with ALMA at Planet-forming Scales Program}

\email{jhc19@mails.tsinghua.edu.cn, weizhu@tsinghua.edu.cn,\\ chrisormel@tsinghua.edu.cn}

\author[0000-0003-2948-5614]{Haochang Jiang (蒋昊昌)}
\affiliation{Department of Astronomy, Tsinghua University, Haidian DS 100084, Beijing, China}

\author[0000-0003-4027-4711]{Wei Zhu (祝伟)}
\affiliation{Department of Astronomy, Tsinghua University, Haidian DS 100084, Beijing, China}

\author[0000-0003-4672-8411]{Chris W. Ormel}
\affiliation{Department of Astronomy, Tsinghua University, Haidian DS 100084, Beijing, China}



\begin{abstract}

Recently, the Molecules with ALMA at Planet-forming Scales (MAPS) ALMA Large Program reported a high number of line-emission substructures coincident with dust rings and gaps in the continuum emission, suggesting a causal link between these axisymmetric line-emission and dust-continuum substructures. To test the robustness of the claimed correlation, we compare the observed spatial overlap fraction in substructures with that from the null hypothesis, in which the overlap is assumed to arise from the random placement of line-emission substructures. Our results reveal that there is no statistically significant evidence for a universal correlation between line-emission and continuum substructures, questioning the frequently made link between continuum rings and pressure bumps. The analysis also clearly identifies outliers. The chemical rings and the dust gaps in MWC 480 appear to be strongly correlated (${>}4\sigma$), and the gaps in the CO isotopologues tend to moderately (${\sim}3\sigma$) correlate with dust rings.

\end{abstract}


\section{Introduction}
\end{CJK*}

In the past decade, the Atacama Large Millimeter/submillimeter Array (ALMA) has provided us with unprecedented imagery on protoplanetary disks starting with HL Tau \citep{ALMAPartnershipEtal2015}. Ubiquitous substructures---including inner cavities, rings, gaps, plateaus, spiral arms, arcs, etc.---are shown in the dust continuum emission in protoplanetary disks surveys \citep[e.g.][]{AndrewsEtal2018,LongEtal2018,CiezaEtal2019,FrancisvanderMarel2020}. Among these substructures, annular rings and gaps are particularly abundant \citep{HuangEtal2018,vanderMarelEtal2019}, whereas nonaxisymmetric substructures are relatively rare \citep{Andrews2020}.

\begin{figure*}
    \centering
    \includegraphics[width=1.98\columnwidth]{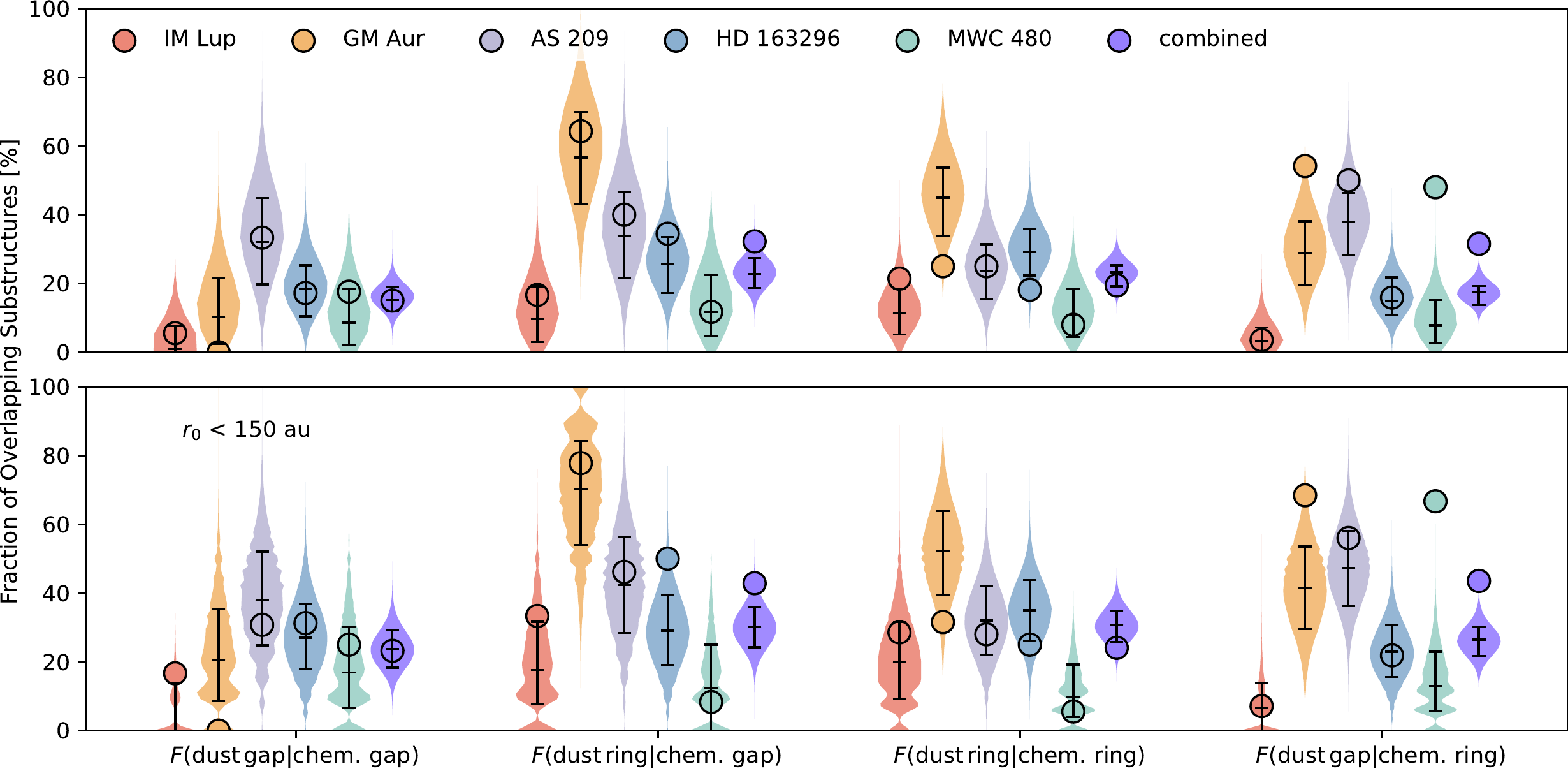}
    \caption{\label{fig:default}\textbf{top:} Fraction of \hjch{chemical substructure}{line-emission substructure}s that spatially overlap with continuum substructures in individual disks and all disks combined. \\
    \textbf{bottom:} The same as the top but only counting \hjch{chemical substructure}{line-emission substructure}s with radial locations $<$150 au.\\
    Circles are observational results. Colored violin plots show the PDF of null hypothesis (\hjch{chemical substructure}{line-emission substructure}s follow random distribution independent of continuum). The error bars show the $[15.9, 50, 84.1]$ percentiles of the probability distribution.}
\end{figure*}

Several formation theories of the dust ring and gap pairs predict that gas and dust substructures should correlate. One such mechanism is the pressure bump model, where the continuum ring is formed by dust trapping inside a local pressure maxima \citep[e.g.][]{PinillaEtal2012,DullemondEtal2018}. Such pressure bumps can be formed by gap-opening planets \citep[e.g.][]{LinPapaloizou1986}, dead zone boundary \citep[e.g.][]{FlockEtal2015}, zonal flow \citep[e.g.][]{BaiStone2014}. Consequently, when the molecular line emission is optically thin the gas surface density enhancement would appear as a chemical ring, leading to an alignment of \hjch{chemical}{line-emission} and continuum substructures. \hjadd{Such correlations between molecular emission lines and continuum rings are observed at the outer edges of inner cavities in many transition disks, which is thought to indicate the presence of giant planets inside the cavities \citep[e.g.][]{vanderMarelEtal2016,vanderMarelEtal2018,DongEtal2017i,BoehlerEtal2017,FedeleEtal2017}. In particular, \citet{FacchiniEtal2021} report this association in the famous two-giant-planet-hosting PDS 70 disk}. Another mechanism \hjadd{that also expects a correlation between gas and dust substructures} is the iceline scenario. Around icelines, evaporation \citep[e.g.][]{ZhangEtal2015} or sintering of dust aggregates \citep[e.g.][]{OkuzumiEtal2016} can contribute to the appearance of a dust ring, while chemical rings and gaps appear when the molecules freeze out beyond the icelines. \hjadd{However, whether the iceline coincides with the dust ring is still a matter of debate \citep{LongEtal2018,HuangEtal2018,vanderMarelEtal2019,ZhangEtal2021i}.} On the other hand, there are also mechanisms of dust ring formation, e.g., the Secular Gravitational Instability \citep{TominagaEtal2020} and the Clumpy Ring Model \citep{JiangOrmel2021}, which do not necessarily require that gas and dust substructures correlate spatially.

Recently, the Molecules with ALMA at Planet-forming Scales (MAPS) \hjch{large program}{ALMA Large Program} allows studies on the gas component of protoplanetary disks at unprecedented resolution and sensitivity \citep{OebergEtal2021}. The MAPS program surveyed approximately 50 molecular lines from 20 different chemical species around five \hjch{disks}{sources}---IM Lup, GM Aur, AS 209, HD 163296, and MWC 480---where dust substructures have been detected in their continuum counterparts \citep{LongEtal2018,AndrewsEtal2018,HuangEtal2020}. In total, more than 200 chemical substructures corresponding to certain molecular lines are identified at high spatial resolutions (7--30 au) \citep{LawEtal2021}. Almost all of these \hjch{chemical substructure}{line-emission substructure}s show signs of an axisymmetric structure \citep{LawEtal2021}. In other words, rings and gaps are also the most abundant morphology in \hjch{chemical substructure}{line-emission substructure}s. These rings and gaps are found at nearly all radii in the observed disks. Many of these \hjch{chemical substructure}{line-emission substructure}s coincide with the dust rings and gaps in the continuum emission, suggesting possible spatial associations between chemical lines and dust continuum \citep{LawEtal2021}. However, statistical tests are needed in order to robustly establish the physical nature of such associations. Because the substructures are so abundant, a random placement of them may also lead to some level of spatial association.

In this work, we quantify the apparent correlation between chemistry and continuum features by means of a null-hypothesis test. We test whether the observed apparent correlation arises solely from randomly distributing the chemical lines. We search for correlations within individual disks (\Se{gbydisk}), and also conduct test on chemical groups (\Se{gbychem}).

\section{Method and Result}\label{sec:method}
A null hypothesis assumes that there is no universal spatial association between dust and \hjch{chemical substructure}{line-emission substructure}s. If the spatial correlation is physical, then the observed overlap fractions should show a significant difference from the overlap fractions from the null hypothesis.

The real \hjch{chemical}{line-emission} distribution varies depending on the disks and the chemical species. To quantify the spatial correlation between the dust and \hjch{chemical substructure}{line-emission substructure}s, we build two groups of statistic tests. We summarize the statistical setup as follows.

\begin{deluxetable*}{lccc|ccc}
\tablenum{1}
\tablecaption{Model Parameters\label{tab:method}}
\tablewidth{0pt}
\tablehead{
\colhead{name} & \multicolumn3c{Rings} & \multicolumn3c{Gaps}
}
\decimalcolnumbers
\startdata
& \# & $r_{\rm in}$ [au] & $r_{\rm out}$ [au] & \# & $r_{\rm in}$ [au] & $r_{\rm out}$ [au] \\
\hline
\multicolumn7l{group by disks} \\
\hline
IM Lup & 28 & 31.5 & 739.6 & 18 & 27.1 & 644.3\\
GM Aur & 24 & 11.8 & 364.3 & 14 & 36.2 & 298.7\\
AS 209 & 28 & 19.3 & 256.8 & 15 & 12.4 & 217.2 \\
HD 163296 & 44 & 12.1 & 418.8 & 29 & 18.0 & 354.8\\
MWC 480 & 25 & 9.2 & 589.1 & 17 & 17.2 & 568.5 \\
combined & 149 & 9.9 & 709.2 & 93 & 13.3 & 602.2 \\
\hline
\multicolumn7l{group by chemical species}\\
\hline
CO & 51 & 9.6 & 729.4 & 36 & 23.5 & 617.4 \\
nitrile & 41 & 15.4 & 418.8 & 25 & 12.7 & 352.1 \\
hydrocarbon & 26 & 22.2 & 408.0 & 13 & 30.8 & 398.8
\enddata
\tablecomments{Model parameters of null hypothesis distributions. (1) sample group name; (2) number of rings; (3) inner boundary of the ring distribution; (4) outer boundary of the ring distribution; (5) number of gaps; (6) inner boundary of the gap distribution; (7) outer boundary of the gap distribution.}
\end{deluxetable*}

\subsection{Quantifying the Spatial Correlation}
We generally follow the approach of \citet{LawEtal2021} in quantifying the correlation between \hjch{chemical}{line-emission} and continuum substructures. We define an overlap as when the radial location of the \hjch{chemical substructure}{line-emission substructure}, $r_{\rm 0, chem}$\footnote{listed in Table 3 of \citet{LawEtal2021}. \hjadd{Following this work, the width of line-emission substructures is not taken into account in this work because widths are affected by beam convolution.}}, falls within the width of the annular continuum substructures
\begin{equation}
    r_{0, \rm dust}-\frac{w_{\rm dust}}{2}
    < r_{0,\rm chem} <
    r_{0, \rm dust}+\frac{w_{\rm dust}}{2}.
\end{equation}
Here $r_{\rm 0, dust}$ and $w_{\rm dust}$ are the radial location and the width of the continuum substructure respectively. \footnote{See Section 3 of \citet{LawEtal2021} for details of characterization of substructures. The locations and widths continuum features are listed in the Table 5 of that work.} The distance between any two dust rings(gaps) is always larger than the sum of their half-width and thus one chemical feature won't fall in two different dust rings(gaps). However, because the dust ring--gap pair could spatially overlap, one chemical feature may fall into one dust ring and one dust gap at the same time.

The overlap fraction is given by
\begin{equation}
    F(\rm A \vert B) = \frac{\rm \#\,of \,feature\,B\,overlap\,with\,feature\,A}{\rm \#\,of\,feature\,B}
\end{equation}
where features A and B are chemical rings and gaps or dust continuum rings and gaps. Four combinations are available. The observed values of $F(\rm dust\vert chem.)$ (hereafter 'chem.' and 'dust' inside formula referring to all four combinations) for individual disks are shown in \fg{default} as open circles, which reproduces the results in the Fig.\,21 of \citet{LawEtal2021}. In addition, we calculate the combined overlap fractions---the sum of overlapped features among 5 disks over the total number of chemical features, which is colored in purple. The ratios range from $0$ to ${\sim}70\%$, depending on the disk and the combination of substructures of chosen. Following \citet{LawEtal2021}, we also calculate the overlap fraction among rings/gaps within 150 au, within which the majority of dust substructures are located. The results are shown as the open circles in the lower panel of \fg{default}.

\begin{deluxetable*}{lcccc|cccc}
\tablenum{2}
\tablecaption{Significance of the Observed Overlap Fractions\label{tab:result}}
\tablewidth{0pt}
\tablehead{
\colhead{name} & \multicolumn4c{among all substructures [$\%\,(\sigma$)]} & \multicolumn4c{for substructures within 150 au [$\%\,(\sigma$)]}
}
\decimalcolnumbers
\startdata
& $F(\rm dust\ gap$ & $F(\rm dust\ ring$ & $F(\rm dust\ ring$ & $F(\rm dust\ gap$ & $F(\rm dust\ gap$ & $F(\rm dust\ ring$ & $F(\rm dust\ ring$ & $F(\rm dust\ gap$ \\
& $\rm \vert chem.\,gap)$ & $\rm \vert chem.\,gap)$ & $\rm \vert chem.\,ring)$ & $\rm \vert chem.\,ring)$ & $\rm \vert chem.\,gap)$ & $\rm \vert chem.\,gap)$ & $\rm \vert chem.\,ring)$ & $\rm \vert chem.\,ring)$ \\
\hline
\multicolumn7l{group by disks} \\
\hline
IM Lup & 76.8 (0.7) & 78.6 (0.8) & 92.6 (1.5) & 58.0 (0.2) & 89.3 (1.3) & 88.6 (1.2) & 77.8 (0.8) & 56.8 (0.2) \\
GM Aur & 10.2 (-1.3) & 71.9 (0.6) & 2.6 (-1.9) & 99.4 (2.5) & 10.1 (-1.3) & 71.0 (0.5) & 4.9 (-1.7) & 98.6 (2.2) \\
AS 209 & 54.2 (0.1) & 68.5 (0.5) & 56.4 (0.2) & 90.9 (1.3) & 31.5 (-0.5) & 62.0 (0.3) & 34.9 (-0.4) & 79.2 (0.8) \\
HD 163296 & 46.0 (-0.1) & 85.4 (1.1) & 5.0 (-1.6) & 53.1 (0.1) & 64.7 (0.4) & 98.1 (2.0) & 12.7 (-1.1) & 45.0 (-0.1) \\
MWC 480 & 83.8 (1.0) & 47.9 (-0.1) & 33.1 (-0.4) & \textbf{$>$99.999 ($>$4)} & 74.8 (0.7) & 29.7 (-0.5) & 23.4 (-0.7) & \textbf{$>$99.999 ($>$4)} \\
combined & 47.5 (-0.1) & 98.3 (2.1) & 15.3 (-1.0) & \textbf{99.995 (3.9)} & 47.0 (-0.1) & 98.5 (2.2) & 7.4 (-1.4) & \textbf{99.992 (3.8)} \\
\hline
\multicolumn7l{group by chemical species}\\
\hline
CO & 66.8 (0.4) & \textbf{99.94 (3.3)} & 26.5 (-0.6) & 99.3 (2.5) & 52.2 (0.1) & 97.7 (2.0) & 5.5 (-1.6) & 98.8 (2.3) \\
nitrile & 45.0 (-0.1) & 23.8 (-0.7) & 5.6 (-1.6) & 98.3 (2.1) & 60.2 (0.3) & 36.2 (-0.4) & 9.7 (-1.3) & 96.4 (1.8) \\
hydrocarbon & 77.9 (0.8) & 49.8 (0.0) & 44.8 (-0.1) & 98.5 (2.2) & 73.0 (0.6) & 67.2 (0.3) & 40.1 (-0.2) & 92.1 (1.4) \\
\hline
& $F(\rm chem.\,gap$ & $F(\rm chem.\,gap$ & $F(\rm chem.\,ring$ & $F(\rm chem.\,ring$ & $F(\rm chem.\,gap$ & $F(\rm chem.\,gap$ & $F(\rm chem.\,ring$ & $F(\rm chem.\,ring$ \\
& $\rm \vert dust\ gap)$ & $\rm \vert dust\ ring)$ & $\rm \vert dust\ ring)$ & $\rm \vert dust\ gap)$ & $\rm \vert dust\ gap)$ & $\rm \vert dust\ ring)$ & $\rm \vert dust\ ring)$ & $\rm \vert dust\ gap)$ \\
\hline
\multicolumn7l{group by chemical species}\\
\hline
CO & 74.8 (0.7) & 98.3 (2.1) & 44.0 (-0.2) & 97.2 (1.8) & 77.4 (0.7) & 93.0 (1.5) & 37.8 (-0.3) & 97.7 (2.0) \\
nitrile & 16.2 (-1.0) & 25.1 (-0.7) & 4.2 (-1.7) & 68.9 (0.5) & 16.7 (-1.0) & 32.9 (-0.4) & 27.6 (-0.6) & 70.4 (0.5) \\
hydrocarbon & 62.6 (0.3) & 46.8 (-0.1) & 46.3 (-0.1) & 93.9 (1.5) & 63.8 (0.4) & 78.7 (0.8) & 84.7 (1.0) & 94.4 (1.6) \\
\enddata
\tablecomments{The percentiles of the observed overlap fractions in randomly generated distributions. The corresponding significances are inside brackets by translation via a standard normal distribution table. (1) sample group name; (2) overlap fraction between chemical gap and continuum gap; (3) overlap fraction between chemical gap and continuum ring; (4) overlap fraction between chemical ring and continuum ring; (5) overlap fraction between chemical ring and continuum gap; (6)-(9) are the same as (2)-(5), but only counting for substructures located within 150 au.}
\end{deluxetable*}

\subsection{Group by disks}\label{sec:gbydisk}
We firstly test the correlation separately for each disk. We construct the substructure distribution of the null hypothesis as follows. For each disk:
\begin{enumerate}
    \item The chemical rings or gaps appear between a certain range $[r_{\rm in}, r_{\rm out}]$. \hjch{Here $r_{\rm in}$ and $r_{\rm out}$ are the inner- and outer-most locations. We derive the boundary values from the detected locations of line-emission substructures in MAPS.\footnote{$r_{\rm in}$ and $r_{\rm out}$ are slightly smaller/larger than locations of observed innermost and outermost features. The offset is analytically calculated to sustain the assumed distribution in (2).}}{Here $r_{\rm in}$ and $r_{\rm out}$ are chosen such that the boundary locations of a finite number of randomly generated substructures better match those detected in observations. As a result, these boundary values are slightly smaller (larger) than the location of the observed innermost (outermost) feature.}\footnote{\hjadd{Specifically, if
    one group of samples has $N$ features in total, the logarithmic distance $L$ between the innermost and outermost feature is  $L=\log{r_{\rm out}^{\rm obs}}-\log{r_{\rm in}^{\rm obs}}$.
    Then, we take
    \begin{equation}
        \log{r_{\rm in}} = \log{r_{\rm in}^{\rm obs}} - \frac{1}{2}\frac{L}{N-1}
    \end{equation}
    and
    \begin{equation}
        \log{r_{\rm out}} = \log{r_{\rm out}^{\rm obs}} + \frac{1}{2}\frac{L}{N-1}
    \end{equation}
    for the inner and outer boundary of the random distribution.
    As the sample size $N$ increases, these numbers will thus get closer to the observed locations. We have verified that our conclusions are unaffected for any choice of the numerical prefactor of the $L/(N-1)$ term in the range between 0 and 1.}} As the rings and gaps appear at different disk radii, gaps are on average closer to the host star than rings. Therefore, we take different $[r_{\rm in}, r_{\rm out}]$ for rings and gaps. \hjch{The values of $r_{\rm in}$ are around $20\pm10$ au for both rings and gaps. While $r_{\rm out}$ ranges from  $\approx\!250\,$au to $\approx\!750\,$au for rings and from  $\approx\!200\,$au to $\approx\!650\,$au for gaps.}{The values we take for each source are listed in \tb{method}.} We tested different values and find that the results are largely unchanged.
    \item The number density of rings or gaps follows the distribution $n(r){\rm d}r \propto r^{-p}{\rm d}r$. We take $p=1$ for all disks. \coch{Such a distribution}{This power-law} approximate\coch{s the}{ly matches the} distribution of the radial locations of \coadd{substructure} shown in Fig.\,17 of \citet{LawEtal2021}. We have also varied the slope between 0.5 and 1.5\hjch{ and our conclusions are unaffected}{, which did not affect our conclusions}.
    \item For each synthetic system, the total numbers of randomly generated chemical rings or gaps are equal to the numbers of rings and gaps observed in the same system by MAPS, respectively. \hjadd{The numbers are listed in \tb{method}.}
\end{enumerate}

We randomly generate substructures following this setup and calculate the overlap fractions. We repeat this process for $10^5$ times for each disk to get the probability density functions (PDFs) of the overlap fraction. The results are shown in \fg{default}.

We evaluate the statistical significance of the observed overlap fractions based on the corresponding PDFs from the simulation. We define the statistical significance as if the probability distribution is of Gaussian shape. Specifically, a detection has a significance of ${<}1\sigma$, ${<}2\sigma$, and ${<}3\sigma$ if the observed value falls within $(15.9,84.1)$, $(2.3, 97.7)$, and $(0.1, 99.9)$ percentiles of the probability distribution derived from the simulation, respectively. \hjch{The derived statistical significances for all combinations of features of all systems are listed}{The percentile values and the corresponding statistical significances of the observed overlap fraction relative to the simulated PDFs are provided} in \tb{result}.

\begin{figure}
    \centering
    \includegraphics[width=0.99\columnwidth]{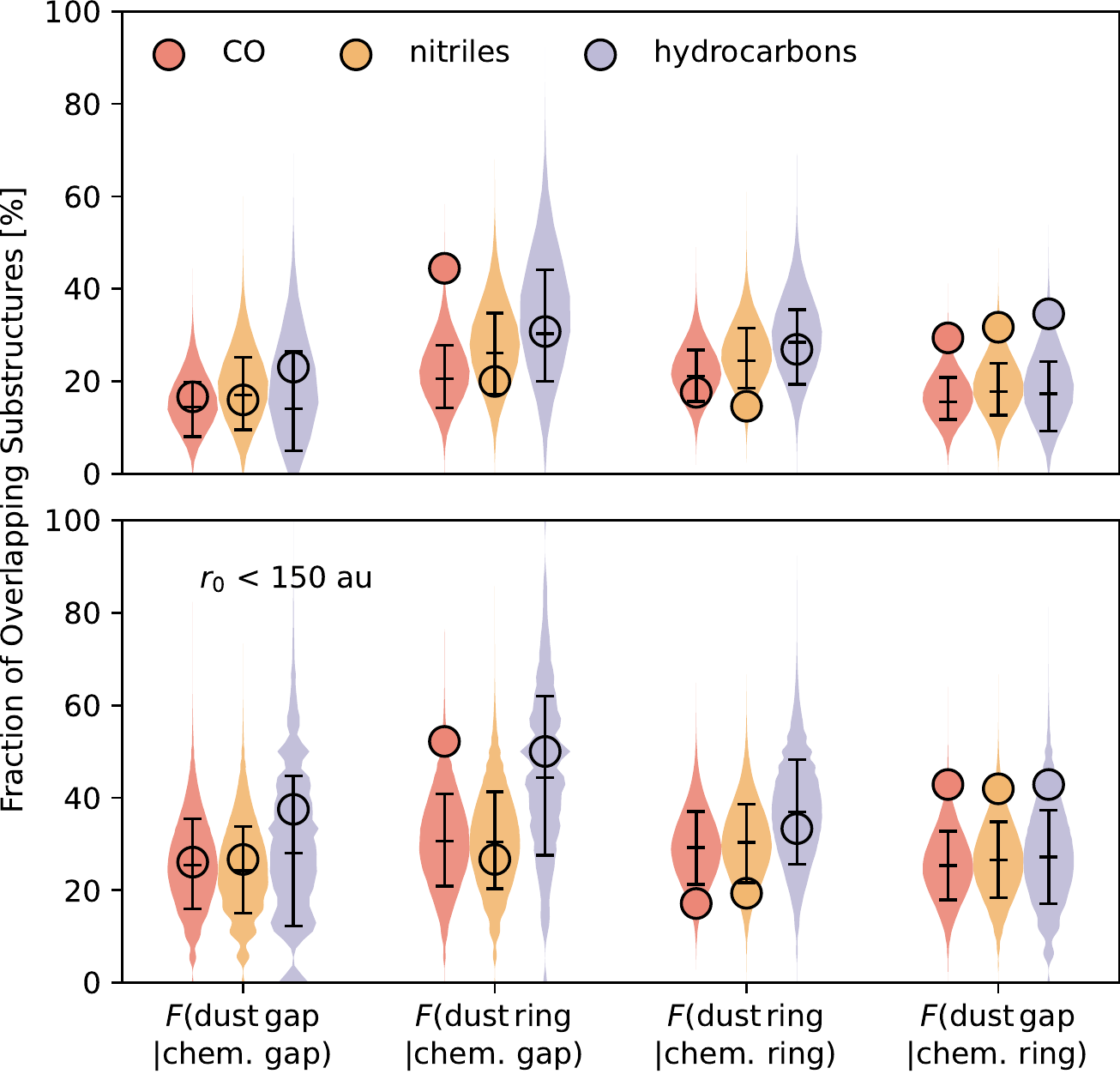}
    \caption{\label{fig:chemgroup1}\textbf{top:} Fraction of \hjch{chemical substructure}{line-emission substructure}s that spatially overlap with continuum substructures in each chemical species group.\\
    \textbf{bottom:} The same as the top but only counting \hjch{chemical substructure}{line-emission substructure}s with radial locations $<$150 au.\\
    Circles are observational results. Colored violin plots show the PDF of null hypothesis (\hjch{chemical substructure}{line-emission substructure}s follow random distribution independent of continuum). The error bars show the $[15.9, 50, 84.1]$ percentiles of the probability distribution.}
\end{figure}

\begin{figure}
    \centering
    \includegraphics[width=0.99\columnwidth]{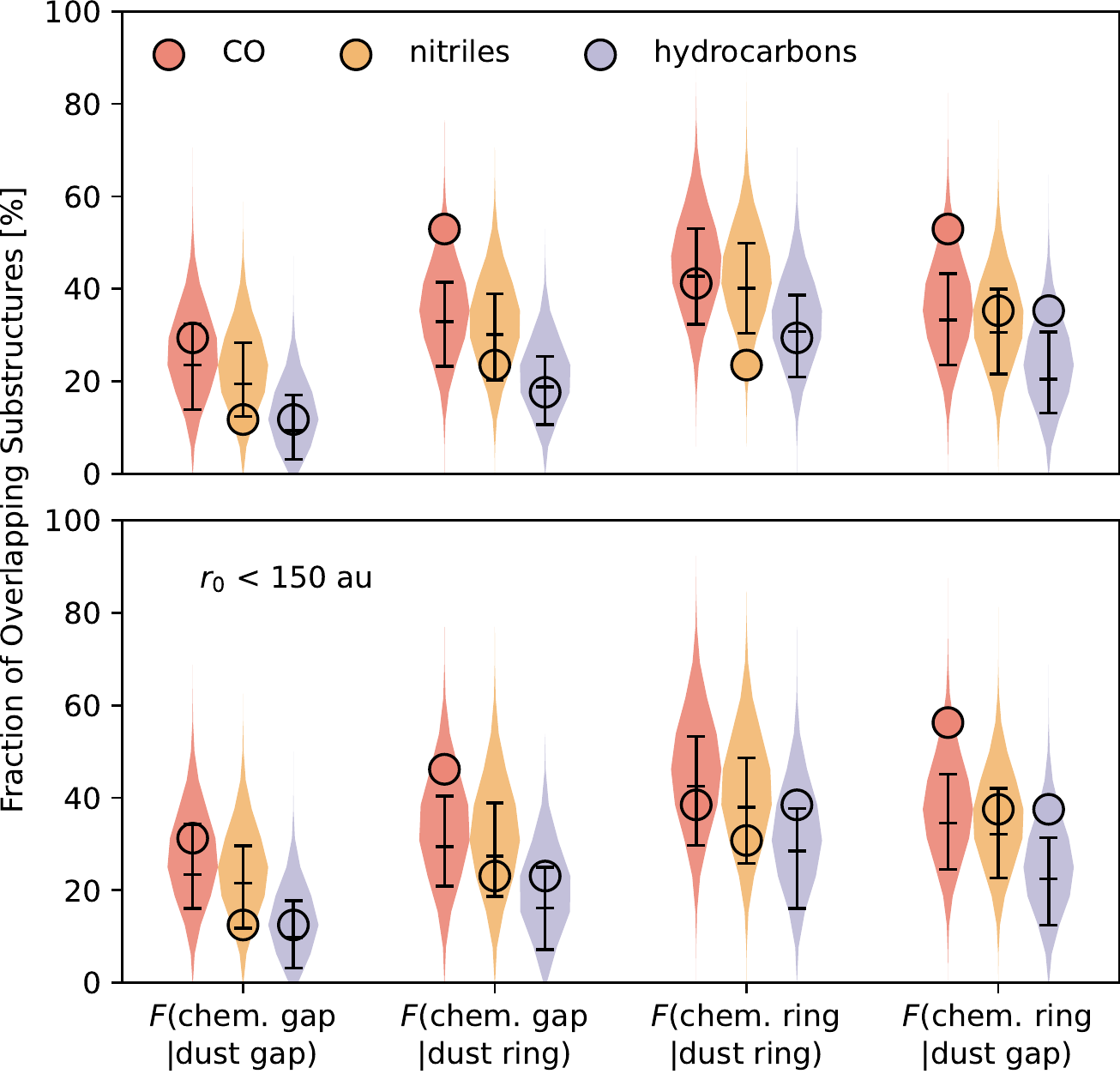}
    \caption{\label{fig:chemgroup2}\textbf{top:} Fraction of continuum substructures that spatially overlap with \hjch{chemical substructure}{line-emission substructure}s in each chemical species group.\\
    \textbf{bottom:} The same as the top but only counting \hjch{chemical substructure}{line-emission substructure}s with radial locations $<$150 au.\\
    Circles are observational results. Colored violin plots show the PDF of null hypothesis (\hjch{chemical substructure}{line-emission substructure}s follow random distribution independent of continuum). The error bars show the $[15.9, 50, 84.1]$ percentiles of the probability distribution.}
\end{figure}

We find that statistically there is no significant evidence for a universal correlation between \hjch{chemical}{line-emission} and continuum substructures. That is, the presence of a \hjch{chemical substructure}{line-emission substructure} does not enhance the probability of finding a dust ring (or gap) in its vicinity. More than half of the observed overlap fractions show statistical significances of ${<}1\sigma$. For individual disks, the observed overlap fractions all fall within the $2\sigma$ ranges of the corresponding PDFs from simulations, except for the overlap fractions, $F(\rm dust\,gap \vert chemical\,ring)$, of MWC 480 (${>}4\sigma$) and GM Aur ($2.5\sigma$). When all disks are combined, there seems to be a statistically significant (${\sim}4\sigma$) correlation between chemical rings and dust gaps. However, such a detection is primarily driven by the single system of MWC 480. Specifically, there are 13 chemical rings located inside the D76 dust gap in MWC 480, making up one third of the (chem.\ ring, dust gap) associations of all five disks. If the MWC 480 system is removed from the analysis, the statistical significance of $F(\rm dust \vert chem.)$ is reduced to below $3\sigma$. Therefore, we conclude that there is no universal correlation between chemical rings and dust gaps.

\subsection{Group by chemical species}\label{sec:gbychem}
Even though \hjch{chemical substructure}{line-emission substructure}s as a whole are distributed randomly for most disks, it is still possible that specific molecular lines may correlate with dust rings or gaps more strongly than others. For example, if a certain molecule X triggers the formation of a continuum ring or gap, it may not easily show up in the previous test where substructures from species X are blended with substructures from all other species. We therefore further investigate the $F(\rm dust\vert chem.)$ by grouping the data with different chemical species. Following \citet{LawEtal2021} (see their Section 3.5), we select and group the chemical species into three groups depending on their chemical similarity---CO isotopologues, nitriles, and hydrocarbons. We collect the chemical features depending on their species group from every disk. Each chemical group is assumed to follow its own random distribution. Therefore, the generation of synthetic systems needs to be modified:
\begin{enumerate}
    \item For each group of chemical species, the spatial range of the randomly placed substructure spans from the inner- and outer-most locations where substructures from this group of species are detected regardless of disks.
    \item For each disk, the number of randomly generated rings/gaps is the same as the number of observed rings/gaps associated with the same group of species. The substructures from different disks are then combined to compute the overlap fraction of the given group of species.
\end{enumerate}
\hjadd{The model parameters are summarized in \tb{method}. }We first calculate $F(\rm dust\vert chem.)$ and show the results in \fg{chemgroup1}. Similar to \fg{default}, the open circles indicate the fractions from observations, the violin plots are used to illustrate the PDFs of the fractions derived from the simulation, and the error bars indicate medians and $1\sigma$ ranges. \hjch{The statistical significance of the observed fractions against the simulation is derived in the same way as in \Se{gbydisk}, and the values are}{The percentiles of observed overlap are} also reported in \tb{result}.

As shown in \fg{chemgroup1} and listed in \tb{result}, the majority of the observed overlap fractions of the three groups of species are statistically insignificant (${<}3\sigma$), regardless of the combinations of substructures. The only exception is the chemical gap and dust ring alignment in CO isotopologues, which has a statistical significance of $3.3\sigma$ (See \se{discussion}). As has been discussed previously, the majority of the \hjch{chemical substructure}{line-emission substructure}s in MWC 480 coincide with the D76 dust gap. This single outlier leads to the marginal detections (${\sim}2\sigma$) of correlations between dust gaps and chemical rings associated with all three groups of species.

By combining the substructures of the same group of species from different disks, we now have a large enough sample to assess the overlap fraction $F(\rm chem.\vert dust)$. This directly informs the fraction of dust substructures associated with certain chemical species. The results are shown in \fg{chemgroup2} and listed in \tb{result}.

The observed overlap fractions are statistically insignificant (${\lesssim}2\sigma$), regardless of the combination of substructures or the group of chemical species. In other words, the presence of a continuum substructure does not significantly enhance, or decrease, the probability of observing a \hjch{chemical substructure}{line-emission substructure} in its vicinity. The results are largely unaffected when only substructures within 150 au are included. This analysis strengthens our conclusion that \hjch{chemical}{line-emission} and dust substructures are generally independent.


\section{Conclusion and Discussion}\label{sec:discussion}
In this work, we revisit the spatial correlation between \hjch{chemical}{line-emission} and continuum substructures in the MAPS survey. We have calculated the observed overlap fractions of different combinations of substructures for individual disks and individual group of chemical species. We assess the statistical significance by comparing the observed fraction with that from synthetic systems that assume no correlation. 
Statistically, the current observational results suggest that the \hjch{chemical substructure}{line-emission substructure}s distribution and dust substructures distribution are overall independent of each other. No significant \hjch{chemical}{line-emission} and continuum substructures correlation is found among the MAPS disks, regardless of \hjch{searching in}{accounting for} the entire disk or \hjch{focusing on}{limiting our analysis to} the inner 150 au. We report our key findings and briefly discuss them as below:
\begin{enumerate}
    \item \coadd{No statistically significant correlation is found between  the  locations  of molecular line  emission  and dust  substructures  in  at  least  four  out  of  the five  MAPS  disks,  regardless  of  which  combination  of  substructure  types  is  considered.   In  particular,  a  positive  correlation  between  dust  rings and chemical rings is absent. Adopting the simplifying assumption that pressure bumps reflect gas surface density maxima, we would naively expect concomitant peaks in the line emission.  Although in reality the actual physical-chemical response to a pressure bump may be more complex such that pressure maxima do not necessarily lead to local chemical emission lines, the lack of a clear continuum--line correlation may alternatively suggest that pressure maxima are perhaps not always the driving mechanism behind continuum rings.
    If this interpretation is right, it would imply that the gas density profile is smooth around continuum and line substructures. Indeed \citet{Alarconetal2021} found that for the AS 209 disk the H$_2$ gas likely follows a smooth \hjadd{surface-density} profile even though there are substructures in certain chemical species.}
    \item Among the five MAPS systems, MWC 480 is a clear outlier with a significant \hjch{detection (${>}4\sigma$) of a correlation}{correlation (${>}4\sigma$)} between chemical rings and dust gaps. This strong correlation is primarily due to the large accumulation of \hjch{chemical substructure}{line-emission substructure}s (${>}13$ rings) at its D76 dust gap.
    Active chemical processing, e.g., gas flow onto the accreting planet(s) \citep[e.g.][]{LiuEtal2019ii,DongEtal2019,TeagueEtal2019}, could be ongoing inside the D76 dust gap in MWC 480.
    \hjadd{\item In addition, a weak (${\sim}2\sigma$) trend of association between line-emission rings and dust gaps appears when focusing on CO isotopologues.
    This meets the scenario demonstrated by some physical-chemical modeling, where the CO emission is increased inside the gaps due to the combination of increased temperature \cite[e.g.][]{vanderMarelEtal2018i,KimEtal2020} and contributions of the backside of the disk \cite[e.g.][]{RabEtal2020}. Specifically, \citet{RabEtal2020} reproduce this association for both CO 2-1 and C$^{18}$O 2-1 emission around the D49 continuum gap in HD 163296. Similar processes may operate in GM Aur to produce the observed trend(${\sim}2.5\sigma$) between dust gaps and CO isotopologues rings.}
    \item We report a moderate (${\sim}3\sigma$) correlation between \hjch{chemical}{line-emission} gaps \hjch{from}{of} CO isotopologues and continuum rings. One possible general explanation has been highlighted by \citet{LawEtal2021}---the overdense regions of dust particles may absorb the optically thick lines and lead to gaps in molecular line observations (see also \citealt{WeaverEtal2018}).
    A naive expectation of this mechanism would be that the same correlation should get stronger with closer-in substructures, because the regions closer to the star should preferentially be optically thicker \citep{BosmanEtal2021}. This is not observed in our results. In fact, the significance of the same correlation decreases when we only count substructures within 150 au.
    \hjadd{Another explanation is that excess cooling of grains inside the dust ring can cause a temperature dip and may lead to a line-emission gap \citep[see Sec. 5.2 of][]{ZhangEtal2021}. Furthermore, optically thick dust rings can absorb the line emission from the back side of the disk \citep{RabEtal2020}. 
    However, \hjch{such a paradigm}{both of these two mechanisms} should apply to other species rather than only CO. Finally, with the concentration of pebbles at the midplane, more efficient physical sequestration of CO onto pebbles and local chemical processing may happen inside continuum rings, which could contribute to such a correlation (See \citealt{KrijtEtal2020} for a more comprehensive study on CO depletion).} More data is needed to confirm this correlation and further analysis and modeling are required to understand its cause.
\end{enumerate}

Given the limitations of current resolution and sample size, one cannot rule out the possibility that some spatial correlations between chemical and dust features may hold for specific disks or specific chemical species.
Better conclusions require higher resolution and a larger sample.
\hjadd{There is increasing evidence that the current chemical inventory in disks may be strongly affected by dust transport, which will lead to significant correlation between chemical components and continuum disk globally \citep[e.g.][]{BosmanBanzatti2019,BanzattiEtal2020,vanderMarelEtal2021i}}. Yet, at present, a universal \hjadd{\textit{local}} continuum--emission-line correlation does not stand. 

\begin{acknowledgments}
    We thank the referee, Nienke van der Marel, for her thoughtful and constructive comments, which significantly improve the quality of this manuscript. H.J. appreciates helpful discussion and comments from Gregory J. Herczeg, Sebastiaan Krijt, Charles J. Law, and Feng Long.
\end{acknowledgments}

\software{
\texttt{Matplotlib} \citep{Hunter2007},
\texttt{Scipy} \citep{VirtanenEtal2020}, 
\texttt{Numpy} \citep{HarrisEtal2020}
}

\bibliography{ads}{}
\bibliographystyle{aasjournal}


\end{document}